\documentclass[twocolumn,prl,superscriptaddress,floatfix]{revtex4}
\usepackage{graphicx}
\newcommand{\kB}{k_{\mathrm{B}}}
\newcommand{\kT}{\kB T}
\newcommand{\rc}{r_c}
\newcommand{\rd}{r_d}
\newcommand{\Lone}{L$_1$}
\newcommand{\Hone}{H$_1$}

\newcommand{\Lalpha}{L$_\alpha$}
\newcommand{\Ltwo}{L$_2$}

\newcommand{\myvector}[1]{{\mathbf{#1}}}
\newcommand{\mytensor}[1]{{\mathbf{#1}}}
\newcommand{\Mtens}{\mytensor{M}}

\newcommand{\nvec}{\myvector{n}}
\newcommand{\rhoC}{\rho_{\mathrm{C}}}

\newcommand{\erf}{\mathrm{erf}}

\begin{document}

\title{Mesoscale simulations of surfactant 
dissolution and mesophase formation}

\author{P. Prinsen}
\affiliation{Unilever Research and Development Port Sunlight,
Bebington, Wirral, CH63 3JW, United Kingdom}
\affiliation{Department of Applied Physics,
Eindhoven University of Technology,
P.O. Box 513, 5600 MB Eindhoven, The Netherlands.}

\author{P. B. Warren}
\affiliation{Unilever Research and Development Port Sunlight,
Bebington, Wirral, CH63 3JW, United Kingdom}

\author{M. A. J. Michels}
\affiliation{Department of Applied Physics,
Eindhoven University of Technology,
P.O. Box 513, 5600 MB Eindhoven, The Netherlands.}

\date{21 April 2002 --- PREPRINT}

\begin{abstract}
The evolution of the contact zone between pure surfactant and solvent
has been studied by mesoscale simulation.  It is found that mesophase
formation becomes diffusion controlled and follows the equilibrium
phase diagram adiabatically almost as soon as individual mesophases
can be identified, corresponding to times in real systems of order
$10\,\mu\mathrm{s}$.
\end{abstract}


\maketitle

When pure surfactant comes into contact with water, mesophases appear
at the interface. This is an important process not only for the
practical use of surfactants, but also from the point of view of
fundamental surfactant phase science. Indeed, contact `flooding' or
penetration scan experiments can yield quantitative information on
mesophases as a function of composition \cite{Laughlin}.  The
phenomenon is invariably \emph{diffusion controlled} in the sense that
the widths of the mesophases follow $t^{1/2}$ growth laws, and
\emph{adiabatic} in the sense that the local composition determines
the mesophase boundaries according to the equilibrium phase diagram
\cite{WB,CMWWRGAL,LaMun}.  But penetration scan experiments are
restricted to observation times of minutes to hours: what happens on
time scales shorter than this?  How early does diffusion control set
in, and how soon can one expect the mesophase boundaries to track the
local composition adiabatically?  Such questions are not just of
scientific interest since time scales of seconds or less are important
in modern processing and the everyday use of surfactants, for instance
determining how rapidly washing powder dissolves.  Experimentally,
this regime is very difficult to access because of the short time
scales and the relatively small amounts of mesophase involved.  To
probe these questions therefore, we have therefore undertaken novel
mesoscale simulations of surfactant dissolution.  We find adiabatic
diffusion control is established remarkably rapidly in our model, on
time scales in which only a few repeat units of the growing mesophases
have appeared, corresponding to times in the real systems of order
$10\,\mu\mathrm{s}$.

The model we have used is a `minimalist' particle-based model of a
binary surfactant / water mixture, based on the dissipative particle
dynamics (DPD) method \cite{GW,JBCKHRW}.  In DPD, the particles are
soft spheres, interacting with pairwise soft potentials of the form $U
= \frac{1}{2} A(1-r/\rc)^2$ ($r < \rc$) where $r$ is the particle
separation, $\rc$ is the range of the interaction, and $A$ the
amplitude.  In the model we have three species of particles: A, B and
C.  The A and B particles are bound together in pairs as dimers with a
fixed separation $\rd$, and represent the surfactants.  The C
particles are monomers representing the solvent (water).  The
different species are distinguished by their interaction amplitudes,
and the trick is to find a set of amplitudes which recover suitable
phase behaviour.  Following earlier work \cite{JBCKHRW}, we use
$A_{\mathrm{AA}} = A_{\mathrm{BB}} = A_{\mathrm{CC}} = 25$,
$A_{\mathrm{AB}} = 30$, $A_{\mathrm{AC}} = 0$, $A_{\mathrm{BC}} = 50$
and $\rd = 0.5$ which gives phase diagram features lying in a
convenient temperature range around $\kT\sim1$ (we fix units by
choosing $m = \rc = 1$ where $m$ is the mass of the particles).

\begin{figure*}
\begin{center}
\includegraphics{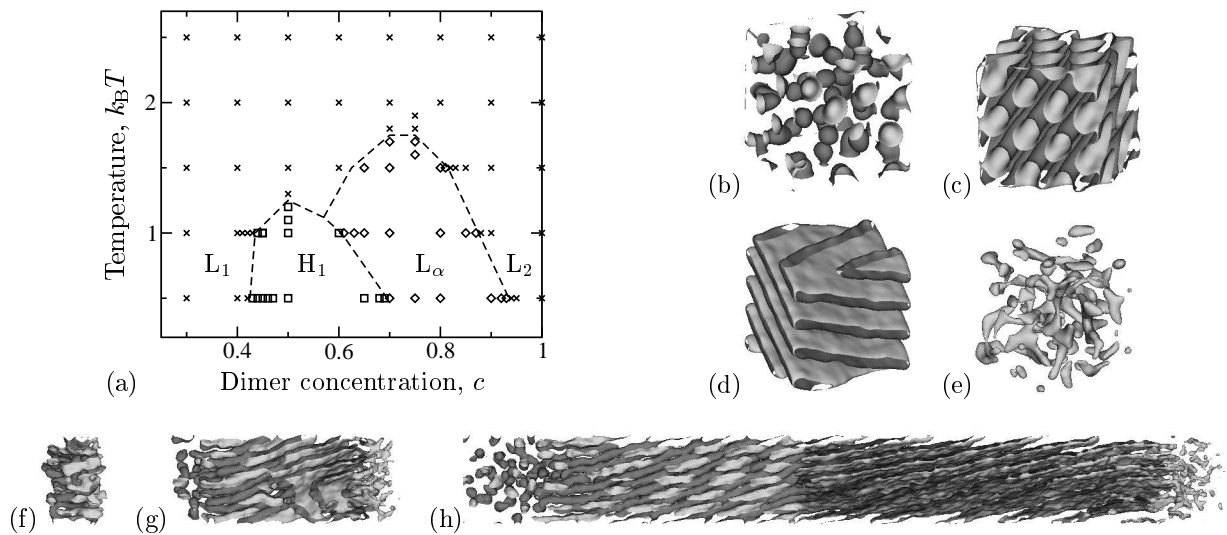}
\end{center}
\caption{(a) Phase diagram for dimer / solvent model in $(c,\kT)$
plane showing points where we find isotropic, hexagonal and lamellar
phases (crosses, squares and diamonds respectively).  The lines are a
guide to the eye.  (b)--(e) Isosurfaces ($\rhoC=1.3$) for equilibrium
phases along $\kT=1$ isotherm, at $c=0.3$, 0.6, 0.7 and 0.9
respectively. (f)--(h) Isosurfaces showing snapshots of surfactant
dissolution simulations, at elapsed times $t=12$, 120 and 1200 DPD
time units from initial contact.\label{fig-one}}
\end{figure*}

Fig.~\ref{fig-one} shows the phase diagram as a function of dimer
concentration and $\kT$, at an overall density $\rho=(2N_{\mathrm{AB}}
+ N_{\mathrm{C}}) / V=6$, where dimer concentration is defined to be
the mole fraction of particles in dimers: $c=2N_{\mathrm{AB}} /
(2N_{\mathrm{AB}} + N_{\mathrm{C}})$.  Similar to many real systems
\cite{LaughlinBook}, micellar (\Lone), hexagonal (\Hone) and lamellar
(\Lalpha) phases are found in order of increasing dimer concentration,
and the interactions are chosen so that on the pure dimer side there
is an isotropic fluid (\Ltwo) phase.  The fact that a certain amount
of solvent is needed to induce the \Lalpha\ phase reflects the real
behaviour of many nonionic surfactants, and is crucial to the contact
simulations below.  We did not find any cubic phases, although these
are not always present in real systems (an extensive study of lattice
models by Larson \cite{surfsim} suggests that cubic phases could be
engineered to appear particularly if the DPD model is elaborated
beyond dimers).  Also, we caution that $\kT$ in the model is not
easily mapped onto a real temperature.  Despite these deficiencies, it
is nevertheless remarkable that such a simple model reproduces the
main features of the phase diagram common to a large number of
surfactants.

To simulate surfactant dissolution and mesophase formation with this
model is now very simple.  We take two simulation boxes, one
containing an equilibrated fluid of pure dimers and the other
containing equilibrated solvent particles.  We place them next to each
other and allow the dimers and solvent particles to interdiffuse.
Just as in the real systems, mesophases start to appear at the
interface over time.  Here we report results for the $\kT=1$ isotherm,
although the $\kT=0.5$ isotherm was also studied with similar results.
We consider simulation boxes of size $10^2\times 100$, with the
concentration gradient along the long axis.  Fig.~\ref{fig-one} shows
some representative simulation snapshots.

Whilst this is straightforward, there are a couple of technical points
to be considered.  Firstly it is important to adjust the densities in
the two boxes so that the pressures are equal.  From separate
simulations, we determined the equation of state for the pure
components, and found that densities $\rho=6.124$ and 5.896 for the
solvent and dimers respectively give a common pressure $p\approx100$.
These densities are within $\approx2$\% of the density $\rho=6$ used
to construct the phase diagram in Fig.~\ref{fig-one}, which
obviates the need to consider constant pressure simulations.  The
pressure matching is important though because it suppresses sound
waves which would spoil the subsequent analysis.  Secondly, if we were
to use conventional periodic boundary conditions, there would be an
unwanted second interface in the system.  We eliminate this by
bounding the simulation box in the long direction by hard reflecting
walls, supplemented by short-range soft repulsive potentials of the
form $U = \frac{1}{2} A_{\mathrm{wall}} (1-z/\rc)^2$ ($z < \rc$),
where $z$ is the distance from the hard wall.  The soft repulsive
force suppresses density oscillations which would otherwise arise due
to the abrupt termination of the particle density.  We find
empirically $A_{\mathrm{wall}}=25$ gives a smoothly vanishing density.
Periodic boundary conditions are retained in the other two directions.

\begin{table}
\caption[?]{Relationship between the ordered eigenvalues $\mu_i$ of
the second moment $\Mtens$ of the isosurface normal distribution
$p(\nvec)$, the nature of $p(\nvec)$, the expected isosurface
geometry, and the expected mesophase; based on Mardia
\cite{MardiaBook}.\label{tab-eval}}
\begin{ruledtabular}
\begin{tabular}{cccc}
Eigenvalues of $\Mtens$ & $p(\nvec)$ & Isosurface & Mesophase\\
\colrule
$\mu_1\approx\mu_2\approx\mu_3\approx1/3$
 & random & spheres/blobs & \Lone\ or \Ltwo \\
$\mu_1\approx0,\,\mu_2\approx\mu_3\approx1/2$
 & planar & cylinders & \Hone \\
$\mu_1\approx\mu_2\approx0,\,\mu_3\approx1$
 & axial & planes & \Lalpha \\
\end{tabular}
\end{ruledtabular}
\end{table}

\begin{figure*}
\begin{center}
\includegraphics{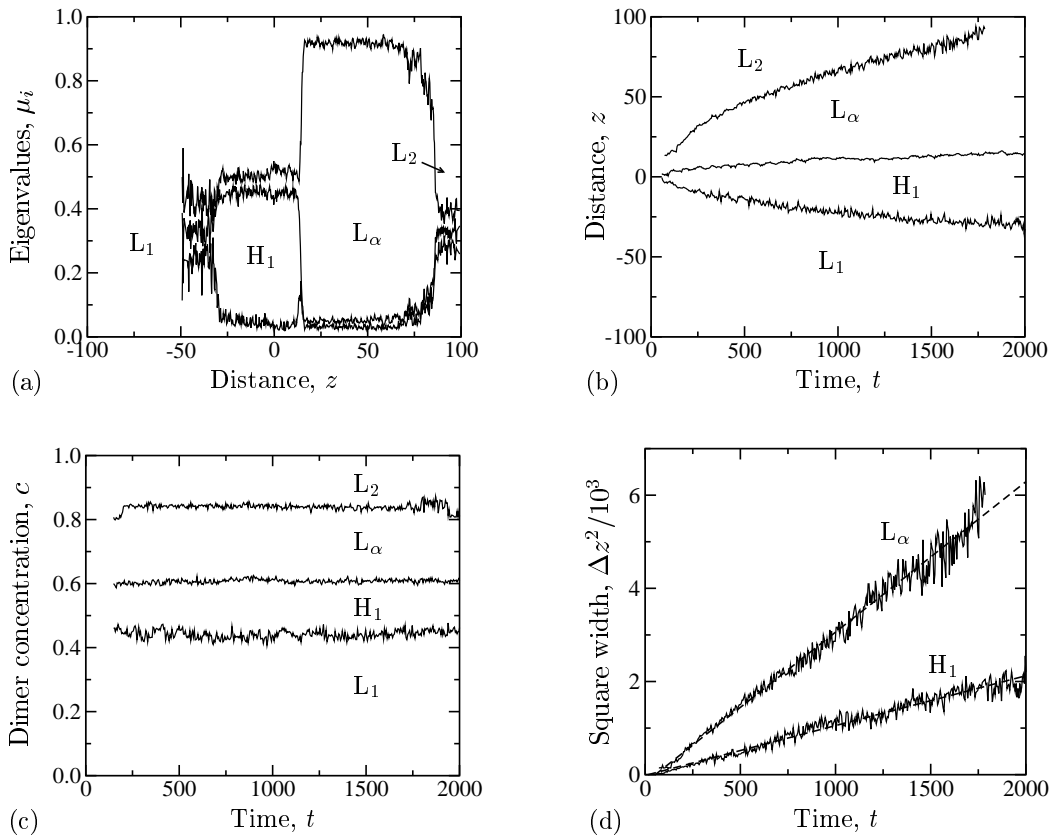}
\end{center}
\caption{(a) Eigenvalues of local order parameter tensor $\Mtens$ at
$t=1200$ DPD time units after initial contact.  (b) Phase boundary
positions as a function of time, as determined from local order
parameter tensor eigenvalues.  (c) Local concentrations at phase
boundaries as a function of time.  (d) Squared widths of middle
mesophases as a function of time.\label{fig-two}}
\end{figure*}

Now we turn to an objective analysis of the simulations to determine
the growth laws.  This is done by introducing a local order parameter
which can distinguish between mesophases as a function of distance
normal to the original contact plane.  Although we examined various
possibilities such as the use of Minkowski functionals \cite{MdR}, an
approach which works well in practice is motivated by the $\rhoC=1.3$
isosurfaces shown in Fig.~\ref{fig-one}.  The different mesophases are
clearly distinguishable to the eye because the isosurface is
predominantly spherical, cylindrical, planar, or fragmented, for the
\Lone, \Hone, \Lalpha\ and \Ltwo\ phases respectively.  This geometric
insight can be made quantitative by constructing the second moment of
the isosurface normal distribution $p(\nvec)$, namely the
symmetric tensor $\Mtens = \int\nvec\nvec\,p(\nvec)\,d\nvec$
\cite{VTK}.  The local geometric nature of the isosurface and hence
the underlying mesophase is reflected in the eigenvalues $\mu_i$ of
$\Mtens$ as laid out in Table~\ref{tab-eval}.  If these eigenvalues
are ranked in order of increasing size, we determined after some
trials that a slice can be classified as \Hone\ if $\mu_1<0.05$ and
$\mu_{2,3}>0.3$, or \Lalpha\ if $\mu_1<0.05$ and $\mu_2<0.15$ (note
$\sum_{i=1}^3\mu_i=1$).

For the present study, we divide the simulation box into slices of
thickness $0.25\,\rc$ parallel to the original contact plane and
determine the eigenvalues of $\Mtens$ for each slice.
Fig.~\ref{fig-two}(a) gives a representative example, showing that the
various mesophases can be clearly distinguished.  The transition
between \Hone\ and \Lalpha\ is particularly sharp as can also be seen
in Fig.~\ref{fig-one}(h).  The above criteria are used to classify
each slice and the positions of the boundaries between mesophases
determined as a function of time.  There is a short incubation period
before the local order parameter can distinguish the different
mesophases, but beyond this point, the boundaries can be reliably
tracked as shown in Fig.~\ref{fig-two}(b).  The key results are
contained in Figs.~\ref{fig-two}(c) and (d), which show respectively
the local composition at the mesophase boundary and the squared width
of the middle two mesophases as a function of time.

Fig.~\ref{fig-two}(c) shows that the local compositions become
established at constant values almost as soon as mesophases can be
distinguished.  Table~\ref{tab-bounds} shows that the \Lone-\Hone\ and
\Hone-\Lalpha\ boundaries in the kinetic simulation lie almost exactly
at the point expected from the equilibrium phase diagram.  The
\Lalpha-\Ltwo\ boundary is displaced from that found in the
equilibrium phase diagram, but it is clear from Fig.~\ref{fig-two}(a)
that this boundary is not completely sharp.  Apart from this boundary
therefore, it appears that the mesophases grow adiabatically almost
from the earliest moments that the local order parameter can
distinguish the growing mesophases.

\begin{table}
\caption[?]{Composition at mesophase boundaries from equilibrium
$\kT=1$ isotherm in Fig.~\ref{fig-one}(a), compared to those
determined from kinetic simulation in Fig.~\ref{fig-two}(c).  Units
are dimer concentration $c$ (the figure in brackets is an
estimate of the error in the final digit).\label{tab-bounds}}
\begin{ruledtabular}
\begin{tabular}{lccc}
Boundary & \Lone-\Hone & \Hone-\Lalpha & \Lalpha-\Ltwo \\
\colrule
Equilibrium & 0.435(5) & 0.605(5) & 0.875(5) \\
Kinetic     & 0.44(1) & 0.61(1) & 0.84(1) \\
\end{tabular}
\end{ruledtabular}
\end{table}

\begin{table}
\caption[?]{Slope of $\Delta z^2$ vs $t$ plot from
Fig.~\ref{fig-two}(d), and effective diffusion coefficient $D$
determined from Eq.~(\ref{eq-deff}) for the two middle mesophases,
compared to self diffusion coefficients for solvent particles and
dimers in the pure components.  Units are DPD units.\label{tab-diff}}
\begin{ruledtabular}
\begin{tabular}{lcccc}
Mesophase & \Hone & \Lalpha & solvent  & dimers \\
\colrule
Slope & 1.07(1) & 3.23(2) & & \\
Effective $D$ & 3.0(5) & 3.2(4) & 0.219(3) & 0.092(2)\\
\end{tabular}
\end{ruledtabular}
\end{table}

Fig.~\ref{fig-two}(d) shows that both mesophases follow a $\Delta
z^2\propto t$ law, where $\Delta z$ is the mesophase width.  This is a
classic verification of diffusion control.  To further examine this,
we fit the slope to what would be expected if the local composition
followed a simple diffusion law:
\begin{equation}
\Delta z^2=4D[\erf^{-1}(2c_2-1)-\erf^{-1}(2c_1-1)]^2\,t,\label{eq-deff}
\end{equation}
where $D$ is an \emph{effective} diffusion coefficient, and $c_{1,2}$
are the fixed compositions at the edges of the mesophase of interest.
The results are shown in Table~\ref{tab-diff}, where for comparison we
have included the diffusion coefficients for the two pure components,
ie dimers in a pure dimer fluid and solvent particles in pure solvent,
determined from separate simulations.  The effective diffusion
coefficients are slightly different in the two phases, similar to the
finding our previous experiments \cite{CMWWRGAL}, although the
difference here is within the statistical errors.  Thus we conclude
that, whilst the diffusion coefficient may be weakly composition- and
mesophase-dependent, the whole process is diffusion controlled from
the earliest times for which mesophases can be distinguished.
Intriguingly the effective diffusion coefficients in the mesophases
are considerably greater than the self diffusion coefficients in the
pure components.  This suggests the existence of a sizeable free
energy driving force enhancing diffusion and is reminiscent of a
detailed study on the \Hone\ phase of the $\mathrm{C_{12}E_6}$ / water
system \cite{SOS}.  Note that the orientation of the \Hone\ and
\Lalpha\ phases in the simulation is such that mutual diffusion occurs
in the `easy' direction: along the hexagonal rods, or parallel to the
lamellar layers (we have also extracted orientation information from
the eigenvectors of $\Mtens$ which will be reported elsewhere).

We can use the effective diffusion coefficients to map elapsed times
in our mesoscale simulation onto real times.  The corresponding
diffusion coefficients in nonionic surfactant systems are typically
$2\times10^{-10}\, \mathrm{m^2\,s^{-2}}$ \cite{CMWWRGAL}.  The
lamellar repeat spacing $L$ can be used as a common measure of length:
$L\approx2.5$ DPD units in the simulation, and typically
$L\approx5\,\mathrm{nm}$ in real systems.  By using $L^2/D$ to scale
time, we conclude one DPD time step is approximately equivalent to
$50\,\mathrm{ns}$.  Thus the duration of the whole simulation (2000
DPD time units) represents more than $0.1\,\mathrm{ms}$ of real time,
equivalent to $10^8$ molecular dynamics time steps.
Fig.~\ref{fig-two} shows that adiabatic diffusion controlled
dissolution sets in after about $200$ DPD time units, equivalent to
about $10\,\mu\mathrm{s}$, which is the origin of the time scale
quoted in the introduction.

Our simulations, whilst addressing very directly intriguing questions
concerning surfactant dissolution, are also interesting because they
are simulations of phase formation kinetics starting from a highly
inhomogeneous state rather than the usual approach which is to quench
a homogeneous system.  To our knowledge, the only comparable study in
the past has been the disappearance of an interface in a vapour-liquid
or binary liquid mixture after an `antiquench' (sudden increase in
temperature) \cite{antiquench}.  It is indeed striking that we do
\emph{not} find any evidence of the delayed appearance of mesophases
due to metastability, unlike that seen experimentally in
temperature-jump experiments \cite{CLLR}, nor do we find any
systematic displacement of the mesophase boundaries by the strong
concentration gradients present at such early times.

Our mesoscale model is also well suited to other fundamental
(meso)phase studies such as temperature quenches \cite{JBCKHRW,CLLR},
the effect of shear on phase boundaries, the exploration of epitaxial
relationships such as between the \Hone\ and \Lalpha\ phases in the
present simulations, amongst many other possibilities.

\bibliography{dissolve}

\end{document}